# Gate Electrostatics and Quantum Capacitance
# of Graphene Nanoribbons

Jing Guo[*], Youngki Yoon, and Yijian Ouyang

Department of Electrical and Computer Engineering

University of Florida, Gainesville, FL, 32611

[*]guoj@ufl.edu

## ABSTRACT

Capacitance-voltage ($C$-$V$) characteristics are important for understanding fundamental electronic structures and device applications of nanomaterials. The $C$-$V$ characteristics of graphene nanoribbons (GNRs) are examined using self-consistent atomistic simulations. The results indicate strong dependence of the GNR $C$-$V$ characteristics on the edge shape. For zigzag edge GNRs, highly non-uniform charge distribution in the transverse direction due to edge states lowers the gate capacitance considerably, and the self-consistent electrostatic potential significantly alters the band structure and carrier velocity. For an armchair edge GNR, the quantum capacitance is a factor of 2 smaller than its corresponding zigzag carbon nanotube, and a multiple gate geometry is less beneficial for transistor applications. Magnetic field results in pronounced oscillations on $C$-$V$ characteristics.

Significant advances have been achieved recently in understanding the fundamental properties and suggesting potential electronics applications of graphene.[1, 2] Graphene is an atomically thin two-dimensional semi-metal with a linear $E$-$k$ relation, which makes it ideal for studying the quantum Hall effect and Dirac Fermion behaviors. It is a promising material for nanoelectronics applications as well. The measured mobility is high (~10,000 cm$^2$/Vs at room temperature[3]), which promises near ballistic transport and ultra-fast switching. Patterning of graphene can lead to semiconducting and metallic graphene nanoribbons (GNR) for electron device and interconnect applications.[4-7] The novel properties of GNRs also make them attractive for non-charge based device applications, such as spintronics.[8]

The capacitance-voltage ($C$-$V$) characteristics are important for understanding the fundamental electronic properties and potential device applications of new nanomaterials. $C$-$V$ measurements of carbon nanotubes (CNTs) have been reported recently, which directly probe van Hove Singularities of CNTs and measure the Luttinger-liquid parameter.[9] The transistor on-current is the average carrier velocity times the charge injected into the channel. The first step to understand transistor performance and to accurately extract mobility is to understand gate electrostatics. Gate electrostatics of GNRs has been studied recently, but the works have been limited to electronic structure calculations[10] and a simple analytical capacitance model[6] of armchair-edge GNRs (A-GNRs) only. The dependences of the GNR $C$-$V$ characteristics on edge shape, gate geometry, self-consistent potential, magnetic field, and their implication for device physics and performance, remain largely unexplored.

In this letter, the $C$-$V$ characteristics of GNRs are simulated by solving the Schrödinger equation in an atomistic basis self-consistently with Poisson equation. The results indicate that the GNR $C$-$V$ characteristics strongly depend on its edge shape. The quantum capacitance of an A-GNR is a factor of 2 smaller than its corresponding zigzag CNT. The small quantum capacitances of A-GNRs have important implications on design of GNR field-effect transistors (FETs). Edge states play a crucial role on the $C$-$V$ characteristics and band structure of a gated zigzag edge GNR (Z-GNR). The highly non-uniform charge distribution in the width direction lowers the total gate capacitance from the gate insulator capacitance considerably. The band structure of a gated Z-GNR can be significantly changed, which leads to a non-zero velocity of the edge states. Application of magnetic field normal to a Z-GNR surface results in pronounced oscillations of the *C-V* characteristics.

Figure 1a shows the cross section of modeled single gate (SG) and double gate (DG) device structures. Both A-GNRs and Z-GNRs are studied. The structures of GNRs are defined in the following ways. The width of an A-GNR is defined by the number of carbon atom dimer lines along the channel, and a Z-GNR is defined by the number of zigzag lines along the channel.[11] The nominal value of the gate insulator thickness is $t_{ins} = 2$ nm and the dielectric constant is $\kappa = 16$.[12] Graphene is amenable to high-$\kappa$ gate and polymer gate insulators due to the lack of the dangling bonds on the surface. A high-$\kappa$ gate insulator is used because it offers better gate electrostatic control. The flat band voltage is set to zero, $V_{fb} = 0$. The simulations were performed at the room temperature, *T*=300 K.

The simulated GNR capacitors are uniform in the direction normal to the cross section in Fig. 1a. The gate is biased at $V_G$ and the GNR is connected to the ground, which can transfer charge to the GNR. The charge density is computed by solving the Schrödinger equation using the atomic Pz-orbital tight binding (TB) model, which has previously been validated by *ab initio* calculations. The edge bond relaxation and spin polarization at edge,[11] which are beyond the scope of the model, can affect the band structures of narrow GNRs, but for the simulated GNR with a width of 5-10 nm, the effects only add small perturbations to the TB results and do not change the qualitative results. The number of electrons at each atomistic site, $N_{p_z}$, can be computed as the diagonal entry of the density matrix,[13]

$$[\rho] = \frac{2}{N}\sum_k f_0([h(k) - \mu I]),$$ (1)

where 2 is the spin degeneracy factor, *k* is the 1D wave vector along the channel, *I* is the identity matrix, $\mu = 0$ is the Fermi level of the GNR (which is grounded), $f_0$ is the Fermi-Dirac function, and *N* is a dummy number of the 1D unit cells. The summation over *k* is performed within the first Brillouin zone numerically. The net number of electrons at each atomistic site is computed as $N_e = N_{p_Z} - N_0$, where $N_0 = 1$ is the number of $P_Z$ electrons per atom when the GNR is charge-neutral. The matrix $h(k)$ is obtained by, [13]

$$h(k) = H_{nn} + \sum_m H_{nm} e^{ikd_{nm}} + U ,$$  (2)

where $H_{nn}$ is the TB matrix of the $n^{th}$ unit cell, $H_{nm}$ is the TB matrix and $d_{nm}$ is the lattice vector between the $n^{th}$ and $m^{th}$ unit cells, and $U$ is a diagonal matrix, whose $i^{th}$ diagonal entry $U_{ii}$ is the on-site electron potential energy at the $i^{th}$ carbon atomistic site. The eigenvalues of $h(k)$ determine the band structure. The effect of the magnetic field is treated by modifying the TB energy between the neighboring $i^{th}$ and $j^{th}$ carbon atoms in the London approximation,[14]

$$t_{ij} = t_0 \exp\left( i \frac{2\pi}{\phi_0} \int_{\vec{r}_i}^{\vec{r}_j} \vec{A}(\vec{r}) \cdot d\vec{r} \right),$$  (3)

where the tight binding parameter is $t_0 \approx 3.0$ eV, $\phi_0$ is the magnetic flux quantum, and $\vec{A}(\vec{r})$ is the vector potential.

After the charge density is computed by solving the Schrödinger equation, the Poisson equation is solved numerically using the finite element method to compute the electrostatic potential $V(\vec{r})$,

$$\nabla \cdot \left[ \kappa(\vec{r}) \nabla V(\vec{r}) \right] = -\frac{\rho(\vec{r})}{\varepsilon_0},$$  (4)

where $\rho(\vec{r})$ is the net charge density due to charging of carbon atoms, and $\kappa(\vec{r})$ is the relative dielectric constant, and $\varepsilon_0$ is the vacuum dielectric constant. The gate voltage $V_G$ is imposed as the boundary condition at the interface between the gate electrode and the gate oxide. For the artificial boundaries that define the numerical simulation region, the Neumann boundary condition, which assumes that the electric field perpendicular to the boundary is zero, is used. We ensure that the simulation region is large enough, so that the simulation results are insensitive to the boundary condition at the artificial boundaries. The potential at the $i^{th}$ atomistic site $V_{S,i}$ is extracted as the potential $V(\vec{r})$, and the on-site electron potential energy is computed as $U_{ii} = -qV_{S,i}$.

The Schrödinger equation and the Poisson equation are solved iteratively. For the $n^{\text{th}}$ iteration:

(i)     The potential energy matrix from the $(n\text{-}1)^{\text{th}}$ step, $U^{n-1}$, is used to calculate the charge density using Eqs. (1) to (3).

(ii)    The Poisson equation, Eq. (4), is numerically solved using the finite element method to compute the potential energy matrix $U^n$.

The iteration continues until the difference between $U^n$ and $U^{n-1}$ is smaller than the error criterion, which is set to 0.1 meV in this work. Once the self-consistency is achieved, the differential gate capacitance is numerically computed,

$$C_G = dQ/dV_G, \tag{5}$$

where $Q$ is the total charge on the gate electrode (which has the same magnitude but an opposite sign as the total net charge of the GNR). Compared to the simple model,[6] the self-consistent atomistic simulation captures the potential variation in the transverse direction of the GNR and its effect on the GNR electronic structure when a non-zero gate voltage is applied.

A simple model is useful for interpreting the numerical simulation results. The gate voltage $V_G$ is the summation of the on-site potential $V_{S,i}$ and the voltage drop over the gate oxide $V_{ox,i}$,

$$V_G = V_{S,i} + V_{ox,i}. \tag{6}$$

If the electrostatic potential on the GNR is uniform, $V_{S,i} = V_S$, the voltage drop over the gate oxide is also uniform, $V_{ox,i} = V_{ox}$, and the following expression can be obtained from Eq. (6),

$$\frac{dV_G}{dQ} = \frac{dV_S}{dQ} + \frac{dV_{ox}}{dQ} \tag{7}$$

If one defines the quantum capacitance as $C_Q = dQ/dV_S$, and the gate insulator capacitance as $C_{ins} = dQ/dV_{ox}$, Eq. (7) becomes

$$\frac{1}{C_G} = \frac{1}{C_Q} + \frac{1}{C_{ins}}, \tag{8}$$

which indicates that the gate capacitance is the serial combination of the quantum capacitance and the gate insulator capacitance. The quantum capacitance,

$$C_Q = \frac{dQ}{dV_S} = q \frac{d \int_{-\infty}^{+\infty} dE \cdot D(E + qV_S) f_0(E - \mu)}{dV_S} = q^2 D_0, \tag{9}$$

where $D(E + qV_S)$ is the density of states (DOS) of the GNR, which is moved by the electrostatic potential, and

$$D_0 = \int_{-\infty}^{+\infty} dE \cdot D(E + qV_S) \left[ -df_0(E)/dE \right] \tag{10}$$

is the weighted average of the DOS near the Fermi level $\mu = 0$.

The above derivation is strictly valid only when $V_{S,i}$ is position-independent. If $V_{S,i}$ is position-dependent, it can be expressed as $V_{S,i} = V_S + \Delta V_{S,i}$, where $V_S$ is the average value of $V_{S,i}$ and $\Delta V_{S,i}$ is the variation from the average value. The average value contributes to $C_Q$ by moving the DOS up and down with reference to the Fermi level, as indicated by Eq. (9). The variation $\Delta V_{S,i}$ contributes to $C_Q$ by changing the band structure and DOS of the GNR. In this case, the definition of $C_{ins}$ breaks down. As discussed in detail later, for an A-GNR, $\Delta V_{S,i}$ is small (compared to the subband spacing) under moderate gate voltages, and the numerical results can be readily interpreted using the simple model. For a Z-GNR, $\Delta V_{S,i}$ is large due to the existence of edge states and the highly non-uniform charge distribution in the transverse direction of the GNR, and the interpretation based on the simple model breaks down. Even for a metallic GNR, the electrostatic potential variation $\Delta V_{S,i}$ can be large in the transverse direction because of the low one-dimensional DOS, the atomically thin and nanometer-wide geometry.

We first examine the dependence of $C_{ins}$ on the GNR width and the gate geometry. Both a single gate (SG) structure and a double gate (DG) structure, as shown in Fig. 1a, are examined.

The SG geometry is easier for fabrication. On the other hand, the DG geometry offers better gate electrostatic control and suppression of electrostatic short channel effects in transistor applications, which has been extensively pursued to extend the scalability and performance of silicon metal-oxide-semiconductor field-effect transistors (MOSFETs). The circles in Fig. 1b shows that the numerically simulated gate insulator capacitances can be well fitted by a simple expression,

$$C_{ins} = N_G \kappa \varepsilon_0 \left( \frac{W}{t_{ins}} + \alpha \right) \tag{11}$$

where $N_G$ is the number of gates (1 for the SG geometry, and 2 for the DG geometry), $\kappa$ is the relative dielectric constant of the gate insulator, $t_{ins}$ is the gate insulator thickness, $W$ is the GNR width, and $\alpha \approx 1$ is a dimensionless fitting parameter. The gate insulator capacitance increases linearly as the GNR width increases because the area of the GNR increases proportionally. The value of $\alpha$ is non-zero due to the electrostatic edge effect. Figure 1b is calculated for the nominal device values, $\kappa = 16$ and $t_{ins} = 2$ nm. We also varied $\kappa$ between 4 and 20, and $t_{ins}$ between 2 nm and 5 nm, and Eq. (11) fits the numerically computed results within 8 % when the GNR width is larger than the insulator thickness, $W > t_{ins}$.

Next we examine the quantum capacitances of A-GNRs when devices operate at quantum capacitance limit ($C_{ins} \rightarrow +\infty$). Figure 1c plots the quantum capacitance $C_Q$ versus the Fermi level $E_F$ with reference to the middle of the GNR band gap $E_m$ for a metallic 41 A-GNR and a semiconducting 42 A-GNR, both of which have widths close to 5 nm. An A-GNR with $n$ dimer lines has the same $E$-$k$ relation as a ($n+1$, 0) zigzag CNT near the Fermi level within the nearest-neighbor tight-binding model.[15] The quantum capacitance curve of a metallic or a semiconducting A-GNR, as shown in Fig.1c, has the same qualitative shape as the DOS of the A-GNR, or its corresponding zigzag CNT, with the Van Hove singularities of the DOS broadened by the thermal energy. An important difference, however, exists between an A-GNR and its corresponding zigzag CNT. The GNR does not have a valley degeneracy factor of 2 because the periodic boundary condition in the circumferential direction of the CNT is replaced by the particle in a box boundary condition in the transverse direction of GNR. The quantum capacitance of an A-GNR, therefore, is a factor of 2 smaller than that of its corresponding zigzag CNT.

The small quantum capacitance of a semiconducting A-GNR has important implications on GNR transistor design and performance. For example, the double gate geometry, which has been extensively pursued for silicon MOSFETs because it increases the charge induced in the channel and on-current by a factor of ~2,[16] can be less helpful for GNRFETs. Figure 1d plots the gate capacitance versus the gate voltage for the SG and the DG GNRs. The DG structure only offers a ~20 % increase of the gate capacitance, which is well below a factor of 2 increase in Si MOSFETs. The reason can be understood as follows. The gate capacitance is the serial combination of $C_{ins}$ and $C_Q$, and the smaller one has a dominant effect. For a typical silicon MOSFET, $C_{ins}$ is smaller and limits the gate capacitance. A DG structure doubles $C_{ins}$ and improves the gate capacitance by ~2. Under a similar gating condition, a GNR can operate at a different limit, in which $C_Q$ is smaller and limits the gate capacitance. The quantum capacitance of a GNR is small because (i) the atomically thin quasi-one-dimensional channel leads to a low DOS and (ii) the quantum confinement boundary condition in the GNR transverse direction further reduces the DOS by a factor of 2 from its corresponding CNT value. When a GNR operates near the quantum capacitance limit, the DG structure doubles $C_{ins}$, but improves the gate capacitance by a much smaller amount, which makes a multiple gate structure less useful in terms of improving GNRFET performance.

The current patterning techniques do not provide atomistic precision of patterning the GNR edge shape, and the GNR edge is likely to be a combination of zigzag and armchair shapes.[4] The effect of the edge shape on the GNR *C-V* characteristics is examined next. We simulated two limit cases, an A-GNR and a Z-GNR, as shown in Fig. 2. Although both of them are metallic and have similar width, Fig. 2a shows that their *C-V* characteristics are drastically different. The gate capacitance of the Z-GNR remains approximately constant in the simulated gate voltage range, -1.5 V < $V_G$ < 1.5 V. In contrast, the gate capacitance of the A-GNR has a constant value in $|V_G| < 0.4$ V due to the constant DOS of the metallic subband, and it reaches local maximum values at $V_G \approx \pm 0.5$ V due to the Van Hove singularities of the lowest semiconducting subband. The features of the A-GNR *C-V* curve can be interpreted by the simple model, in which the gate capacitance is the serial combination of the gate insulator capacitance and the quantum capacitance.

The simple model, however, can not be extended to understand the $C$-$V$ characteristics of the Z-GNR, which has the following two major features. First, the gate capacitance of the Z-GNR remains approximately constant in the simulated $V_G$ range, although the DOS of the Z-GNR has a sharp singularity near $E$=0 due to edge states (as shown in Fig. 2b). Second, the gate capacitance value is considerably smaller than $C_{ins}$≈4.9 pF/cm (as shown in Fig. 1b), and it is even smaller than that of the 41 A-GNR with a similar width when $\left| V_G \right| > 0.5$ V. These phenomena can be explained by the fact the charge distribution and the electrostatic potential are highly non-uniform in the transverse direction of the Z-GNR. The DOS singularity at $E$=0 (as shown in Fig. 2b) is contributed by the edge states, which only results in charge distribution at the GNR edges but not in the middle of the GNR.[11] The gate essentially modulates the charge on two zigzag atomic chains at the edges of the GNR, and the zigzag atomic chains in the middle of the GNR are nearly insulating and the gate can hardly induce charge on them. As a result, the charge distribution in the transverse direction is highly non-uniform, as shown by the dashed line in Fig. 3a, which plots the number of net electrons and the electron potential energy along the transverse direction of the Z-GNR at $V_G$=1.5 V. Because the insulator capacitance $C_{ins}$≈4.9 pF/cm (in Fig. 1) is computed by assuming that the electrostatic potential is uniform on the GNR, this concept breaks down for the Z-GNR. The simulated gate capacitance is small and limited by the electrostatic gating geometry, in which the gate induces the charge only into two zigzag atomic chairs at the edges. Because the gate capacitance is limited by the electrostatic gating geometry, the singularity of the Z-GNR DOS at $E$=0 is not pronounced in the simulated $C$-$V$ characteristics.

Another important effect of the edge states is that the band structure and DOS of a Z-GNR can be significantly altered by the self-consistent potential even at modest gate voltages. A previous study has shown that for A-GNRs, the perturbation of self-consistent potential on the GNR band structure is small.[10] In contrast, the effect is much larger for Z-GNRs due to the existence of edge states. Figure 3a shows that the variation of the electron potential energy in the transverse direction of a 5 nm wide Z-GNR is as large as ~500 meV at $V_G$=1.5 V. The charge density at the edges is large due to the edge states, which produces a large self-consistent potential that prevents the applied gate voltage to lower the electron potential energy at the edges. In contrast, the charge density at the middle of the GNR is small, which results in a small self-consistent potential. The positive gate bias can lower the electron potential much more effectively. As a result, the potential variation on the Z-GNR is large, as shown by the solid line in Fig. 3a. Even under modest gate bias, the band structure can be changed by the large variation of the potential in the transverse direction, as shown in Fig. 3b and Fig. 3c. As a result, the group velocity of the edge states changes from 0 to a finite value, which has an important effect on the speed of signal propagation along the Z-GNR, and thereby, the intrinsic speed of GNR devices and interconnects with zigzag edges.

Magneto-transport in graphene is a topic of strong interest.[1, 2] The effect of magnetic field on the *C-V* characteristics of gated GNRs is examined next. Figure 4 plots the *C-V* characteristics of a Z-GNR with a width of 9.9 nm without and in the presence of a magnetic field, $B$=15.6 T, normal to the GNR surface. The results show that the magnetic field further increases the gate capacitance near $V_G = 0$ V, and leads to more pronounced oscillations on the *C-V* characteristics. The reason is that the magnetic field results in the formation of Landau energy levels,[1, 2] which leads to the oscillations of the LDOS in the middle of the GNR. Because the lowest Landau level forms close to $E$=0, the capacitance around $V_G = 0$ V increases. The effect of magnetic field on causing the oscillations of the gate capacitance is preserved even when $V_G$ increases and the self-consistent potential becomes important. The peak values in the gate capacitance curve indicate that the Fermi level of the GNR channel aligns with a Landau level, which significantly increases the LDOS in the middle of the GNR. The *C-V* measurement, therefore, could be useful for probing the formation of Landau levels in gated GNRs.

The *C-V* characteristics of GNRs are significantly different from CNTs due to the existence of the edges. For an A-GNR, the edges result in a different boundary condition and a factor of 2 smaller quantum capacitances than its corresponding zigzag CNT. For a Z-GNR, the edge states result in highly non-uniform charge distribution in the transverse direction, which significantly lowers the gate capacitance value. The electrostatic potential variation in the transverse direction of the GNR is large and the simple definition of the gate insulator capacitance breaks down. The DOS singularity due to the edge states of a Z-GNR does not result in a peak value in gate capacitance. The potential variation also significantly changes the band structure of a gated Z-GNR. The unique features of the GNR *C-V* characteristics can be tested experimentally by using the recent advance on measuring the *C-V* characteristics of 1D nanomaterials.[9] In addition, understanding GNR gate electrostatics and quantum capacitance is a necessary step for understanding the device physics and assessing the performance of GNRFETs.

Acknowledgement

The authors would like to thank Seokmin Hong for extensive discussions.

# REFERENCES

1. Novoselov, K. S.; Geim, A. K.; Morozov, S. V.; Jiang, D.; Zhang, Y.; Dubonos, S. V.; Grigorieva, I. V.; Firsov, A. A. *Science* **2004,** 306, (5296), 666-669.
2. Zhang, Y. B.; Tan, Y. W.; Stormer, H. L.; Kim, P. *Nature* **2005,** 438, (7065), 201-204.
3. Berger, C.; Song, Z. M.; Li, X. B.; Wu, X. S.; Brown, N.; Naud, C.; Mayo, D.; Li, T. B.; Hass, J.; Marchenkov, A. N.; Conrad, E. H.; First, P. N.; de Heer, W. A. *Science* **2006,** 312, (5777), 1191-1196.
4. Chen, Z. H.; Lin, Y. M.; Rooks, M.; Avouris, P. *cond-mat/0701599* **2007**.
5. Obradovic, B.; Kotlyar, R.; Heinz, F.; Matagne, P.; Rakshit, T.; Giles, M. D.; Stettler, M. A.; Nikonov, D. E. *Applied Physics Letters* **2006,** 88, (14), 142102.
6. Ouyang, Y.; Yoon, Y.; Fodor, J. K.; Guo, J. *Applied Physics Letters* **2006,** 89, (20), 203107.
7. Areshkin, D. A.; Gunlycke, D.; White, C. T. *Nano Letters* **2007,** 7, 204-210.
8. Son, Y. W.; Cohen, M. L.; Louie, S. G. *Nature* **2006,** 444, (7117), 347-349.
9. Ilani, S.; Donev, L. A. K.; Kindermann, M.; McEuen, P. L. *Nature Physics* **2006,** 2, (10), 687-691.
10. Fernandez-Rossier, J.; Palacios, J.; Brey, L. *cond-mat/0702473* **2007**.
11. Nakada, K.; Fujita, M.; Dresselhaus, G.; Dresselhaus, M. S. *Physical Review B* **1996,** 54, (24), 17954-17961.
12. Javey, A.; Guo, J.; Farmer, D. B.; Wang, Q.; Wang, D. W.; Gordon, R. G.; Lundstrom, M.; Dai, H. J. *Nano Letters* **2004,** 4, (3), 447-450.
13. Datta, S., *Quantum transport : atom to transistor*. Cambridge University Press: Cambridge, UK ; New York, 2005; p xiv, 404 p.
14. Zhang, Y.; Yu, G. L.; Dong, J. M. *Physical Review B* **2006,** 73, (20), 205419.
15. White, C. T.; Li, J. W.; Gunlycke, D.; Mintmire, J. W. *Nano Letters* **2007,** 7, (3), 825-830.
16. ITRS, *International Technology Roadmap for Semiconductors, www.itrs.net*.

**List of Figures**

Fig. 1   (a) The single gate (SG) structure (left) and the double gate (DG) structure (right). The nominal value of the gate insulator thickness is $t_{ins} = 2$ nm and the relative dielectric constant is $\kappa = 16$. (b) The numerically simulated *gate insulator capacitance* versus GNR width for the SG structure (the solid line) and the DG structure (the dashed line). The circles are the linear fitting to the simulated capacitance. (c) The quantum capacitance versus the Fermi level for a 41 A-GNR, which is metallic and has a width of 4.9 nm (the dashed line), and for a 42 A-GNR, which has a width of 5.0nm and a band gap of 0.25 eV (the solid line). (d) The gate capacitance versus the gate voltage for the SG (the solid line) and the DG (the dashed line) structures of the 42 A-GNR. The temperature is $T$=300 K.

Fig. 2   *Zigzag edge GNR* (a) The *C-V* characteristics of a 24 Z-GNR, which has a width of 5.0 nm (the solid line), compared to a 41 A-GNR, which has a width of 4.9 nm (the dashed line). The single back gate geometry as shown in Fig. 1a is used. The temperature $T$=300 K. (b) DOS of the 24 Z-GNR (the solid line) and the 41 A-GNR (the dashed line) at $V_G$=0, which results in a zero self-consistent electrostatic potential.

Fig. 3   *Effect of edge states* (a) The electron potential energy (the left axis, solid line) and the net number of electrons per atom (the right axis, dashed line) versus the dimer index of an armchair chain in the transverse direction of the simulated 24 Z-GNR, which has 48 carbon atoms. The gate voltage $V_G$=1.5 V. (b) The *E-k* relation at $V_G$=1.5 V, compared to (c) the *E-k* relation at $V_G$=0.

Fig. 4 *Effect of magnetic field* The *C-V* characteristics of a 47 Z-GNR, which has a width of 9.9nm, in the absence of magnetic field (the solid line) and in the presence of magnetic field $B = 15.6$ T (the dashed line). Single back gate geometry in Fig. 1a is used.

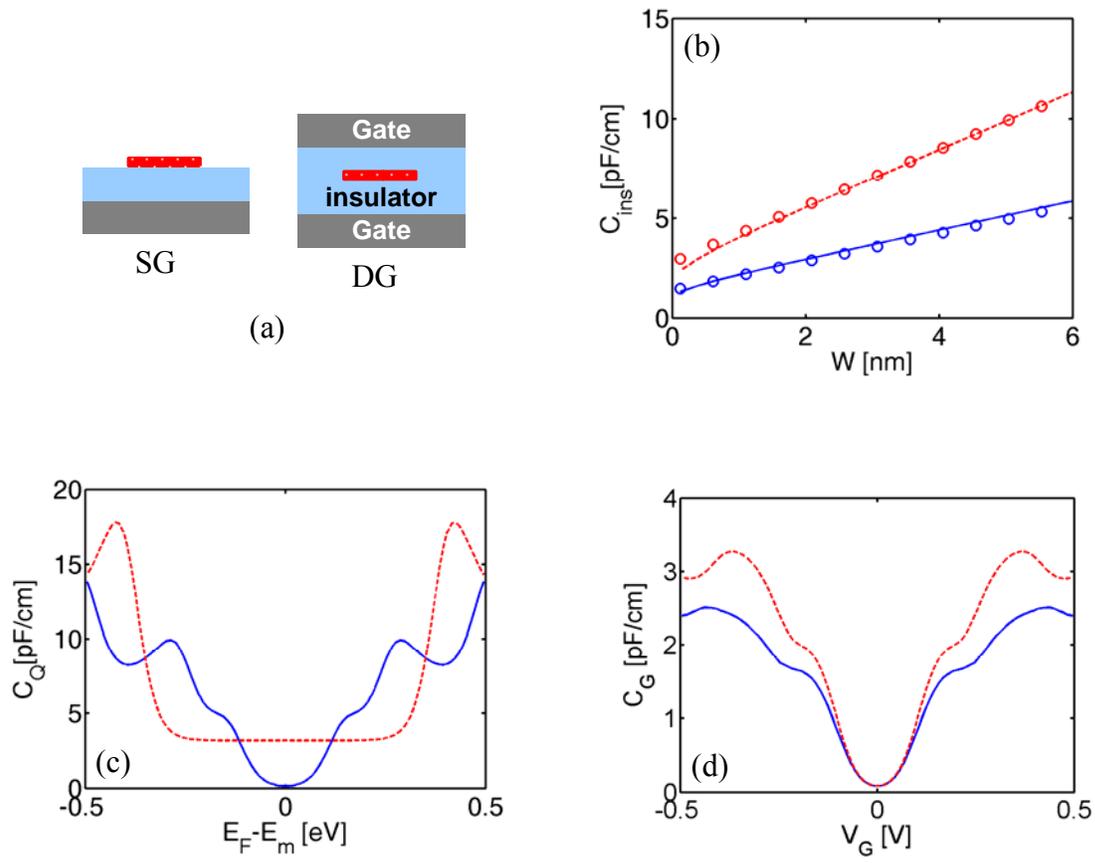

FIGURE 1

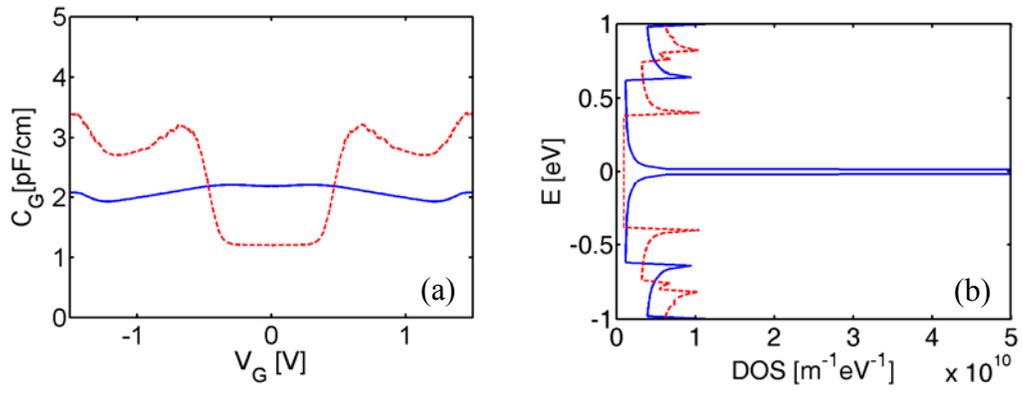

FIGURE 2

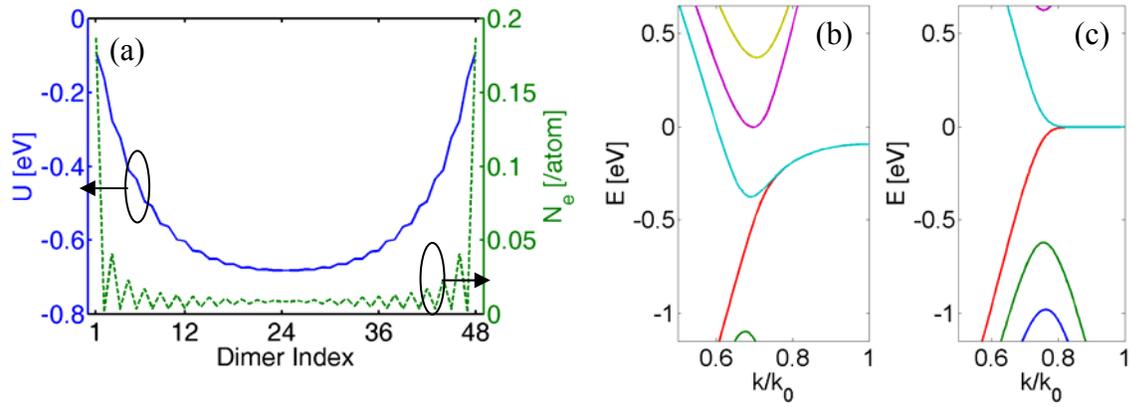

FIGURE 3

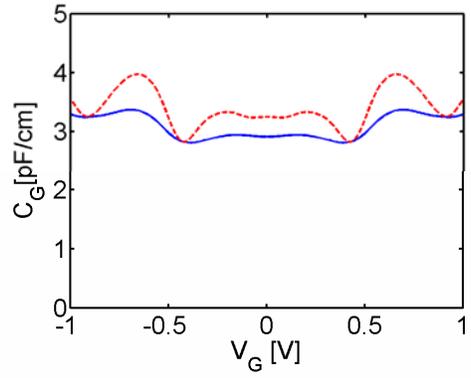

FIGURE 4